\documentclass[12pt]{article}
\usepackage{graphicx}
\usepackage{url}

\title{Memories of Murray and the Quark Model\footnote{
Talk presented at the ``Conference in Honor of Murray Gell-Mann's 80th Birthday,'' Nanyang Technical University, Singapore, February 24, 2010.} }

\author{George Zweig\\
26-169, Research Laboratory of Electronics\\
Massachusetts Institute of Technology \\
77 Massachusetts Ave.\\
Cambridge, MA 02139-4307, USA\\
zweig@mit.edu
}
\date{June 30, 2010}

\begin{document}
\maketitle
\begin{abstract}
\noindent Life at Caltech with Murray Gell-Mann in the early 1960's is remembered. Our different paths to quarks, leading to different views of their reality, are described. 
\end{abstract}

\pagenumbering{arabic}

{\it Prologue:} In 1964 Dan Kevles arrived at Caltech from Princeton as a young assistant professor of history, specializing in the history of science. As an undergraduate he had majored in physics. Shortly after his arrival I barged  into his office, told him that Elementary Particle Physics was in great flux, tremendously exciting; history was in the making, just waiting for him to record. And much of it involved Richard Feynman and Murray Gell-Mann, whose offices were just 600 feet away!

My excitement was not contagious. Dan lectured me, saying that no one can recognize what is historically important while it is happening. One must wait many years to understand the historical significance of events. What he could have added is that it is convenient for historical figures to be unavailable to contradict historians who document their actions, and sometimes even their motives.\footnote{In Dan's defense, when Henry Kissinger asked China's Premier Zhou Enlai to assess the 1789 French revolution, Zhou Enlai is reported to have replied, ``It is too early to say.''}

Well, I'm going to risk it. Today I'm going to tell you  about the Murray Gell-Mann I saw in action, and a little bit about the history of the quark model. Not only is Murray alive and well, he's in the audience, and will keep me honest. So let's begin.

{\it Early influences:} Murray, a belated ``Happy Birthday!''  I have learned a lot from you, and for that I am truly grateful. We go way back, even further than you realize. In the summer of 1957, after a hard day's work as a counselor at a day camp for children, I came across an article you coauthored in {\it Scientific American}, which said of elementary particles~\cite{MGMsciAm1}:

\begin{quote}
``At present our level of understanding is about that of Mendeleyev, who discovered only that certain regularities in the properties of the elements existed. What we aim for is the kind of understanding achieved by Pauli, whose exclusion principle showed why these regularities were there, and by the inventors of quantum mechanics, who made possible exact and detailed predictions about atomic systems.''
\end{quote}
This article appeared just three months before  Sputnik, when it still wasn't fashionable to do physics.
At the time I was just starting my junior year at the University of Michigan as a math major, but was thinking of switching to physics when going to grad school. Here was a big green light saying: ``Go!''

In my senior year I went in to see my quantum mechanics professor P.V.C. Hough for advice on graduate schools. This was the Hough who would become the Hough of the Hough-Powell bubble chamber digitizer, and the Hough transform in image processing. His comment: ``Bethe is at Cornell, where I come from, but he's getting old. There are a couple of young guys at Caltech, Feynman and Gell-Mann, why don't you go there.'' And I did.

{\it Life at Caltech, the first 3 years: } It was wonderful to be at Caltech in the very early 60's. Carl Anderson was the avuncular chairman of the physics department. The theory graduate students included Hung Cheng, Sidney Coleman, Roger Dashen, Jim Hartle, and Ken Wilson, just to name some. All but Roger were Murray's graduate students. Shelly Glashow and Rudy M\"{o}ssbauer were postdocs, and Yuval Ne'eman and J.J. Sakurai were visitors. And then, of course, there were Murray and Richard Feynman. If that wasn't enough, you could always go across campus and talk with ex-particle-physicist Max Delbr\"{u}ck, who had invented molecular biology, or Linus Pauling, a phenomenologist par excellence.

Money was pouring into particle physics, helped now by Sputnik. Pictures from bubble and spark chambers were just beginning to provide an enormous wealth of information. I still remember driving across LA to my first APS meeting at UCLA. In a cavernous dark half-empty auditorium three speakers, Bogdan Magli\'{c}, Bill Walker, and Harold Ticho showed slides demonstrating the existence of the first meson resonances, the $\omega$, $\rho$, and $K^{*}$.  APS meetings seemed pretty interesting!

Shortly thereafter, Murray and Yuval Ne'eman independently proposed that these, and other hadronic resonances, be classified according to the representations of SU(3),  trumping Lee and Yang who continued to use the representations of $G_2$~\cite{LeeYang}. But this is getting ahead of our story.

After my first academic year at Caltech I asked Bob Christy, one of my professors, if I could do theoretical research with him over the summer. In a very disdainful way he replied, ``You know nothing. Why don't you go over to the Synchrotron and learn experimental physics. If you do become a theorist later you won't have time to learn what experimental physics is all about.'' In retrospect, this was great advice.

At the Synchrotron, Alvin Tollestrup was testing his ``fast electronics,'' which would be used to study the nonleptonic decay $K^{+}\rightarrow \pi^{+}+\pi^{0}+\gamma$ at the Bevatron in Berkeley. This {\it K} particle had other uses.   After talking to Alvin, I proposed looking for the violation of time-reversal symmetry in leptonic {\it K}-decay, piggybacking on Alvin's experiment. This was to be my thesis problem. Alvin suggested that I talk to Murray to gain a better understanding of the $\Delta I = 1/2$ rule in nonleptonic {\it K}-decay, which Alvin's experiment was designed to illuminate.

At this point I remember only one of my meetings with Murray. I had worked out a dynamical mechanism for the suppression of leptonic {\it K} decay, which allowed me to predict angular distributions. My first theoretical result! I walked happily into Murray's office, handing him two pieces of paper, one a xerox copy of the published experimental results, and the other the corresponding theoretical angular distributions, which were in good agreement with experiment. Murray looked at the two pieces of paper, looked at me, and said ``In our field it is customary to put theory and experiment on the same piece of paper.'' I was mortified, but the lesson was valuable.

Because I was a graduate student, I got off lightly in my interactions with the faculty. Not so for all. Fred Zachariasen, who had initially suggested Alvin's {\it K}-decay experiment, invited one of his collaborators, Marshall Baker, to give a seminar about Marshall's recent work on {\it K}-decay. Particle physics seminars took place every Tuesday at 2 o'clock in a very small classroom.
As usual, Feynman and Murray sit front row center.  Lesser luminaries, postdocs, and graduate students sit in the rows behind them. Murray is wearing his tweed sports coat with tie, while Feynman, dressed more like a graduate student, impatiently taps the floor with his hush puppy shoes. Both of them look oddly out of place, squeezed into drop-leaf chairs, with their paddles out, meant for undergraduates. As Marshall begins, Murray reaches down to his side, picks up a folded newspaper from the floor, unfolds it, snaps it open at eye level, and proceeds to read right in front of Marshall, who is only a yard away. After about a minute, Feynman, who doesn't pay much attention to other people's work,  leans over to Murray and asks in his best Far Rockaway accent ``Is this guy smart?'' Feynman's voice is hushed, but loud enough so that everyone in the room, including the speaker, hears the question. This is not the first time the seminar attendees have witnessed these two in action. They know that if Murray's head nods up and down behind the paper, Feynman will ask questions. If his head rocks back and forth, Feynman won't waste time with questions. This time Murray nods up and down, answering the question for everyone except the speaker. What the seminar attendees didn't know is that Marshall stutters when stressed. Feynman starts questioning, Marshall starts stuttering; the more questions, the longer the stutter. With Feynman's final question, Marshall's stutter goes into an infinite loop, Feynman slams the palm of his hand down on the paddle of his drop-leaf chair, shouts ``Goddamn it! I can't get a straight answer out of this guy,'' and storms out of the classroom, leaving Marshall in full stutter.

The next day I happened to walk by Murray's office. The door was open, and I overheard Fred animatedly asking Murray to give Marshall a \$100 honorarium as partial compensation for Feynman's atrocious behavior. Murray seemed sympathetic, but noncommittal.\footnote{Speakers at Caltech theory seminars never got an honorarium. Murray no longer remembers if this tradition was broken in Marshall's case.}

I won't describe the next two years of 18-hour days of classes and experimental work. When the smoke cleared, I couldn't find any evidence for the violation of time-reversal symmetry. Faced with the prospect of another two years determining the value of an upper bound, I punted and went to Mexico for a month. Upon returning, I switched to theory, and asked Murray to be my thesis advisor. Despite what you might think from my previous remarks, Murray had been very kind to me, almost fatherly, so he was a natural choice. But Murray said no! He was going to the East Coast on sabbatical, but he ``would talk to Dick.'' 

When I went in rather timidly to ask Feynman if he would be my thesis advisor, he responded: ``Murray says you're OK, so you must be OK.'' And then I remembered Murray's nodding up and down at Marshall Baker's seminar. After telling me about life with his thesis advisor, Johnny Wheeler, Feynman said that he wanted to see me from 1:30 in the afternoon till tea time (4:15) every Thursday. I prepared frantically for each meeting, never presenting the same topic twice. This went on for the entire academic year.

\begin{figure}
\centering
\includegraphics[scale=0.14]{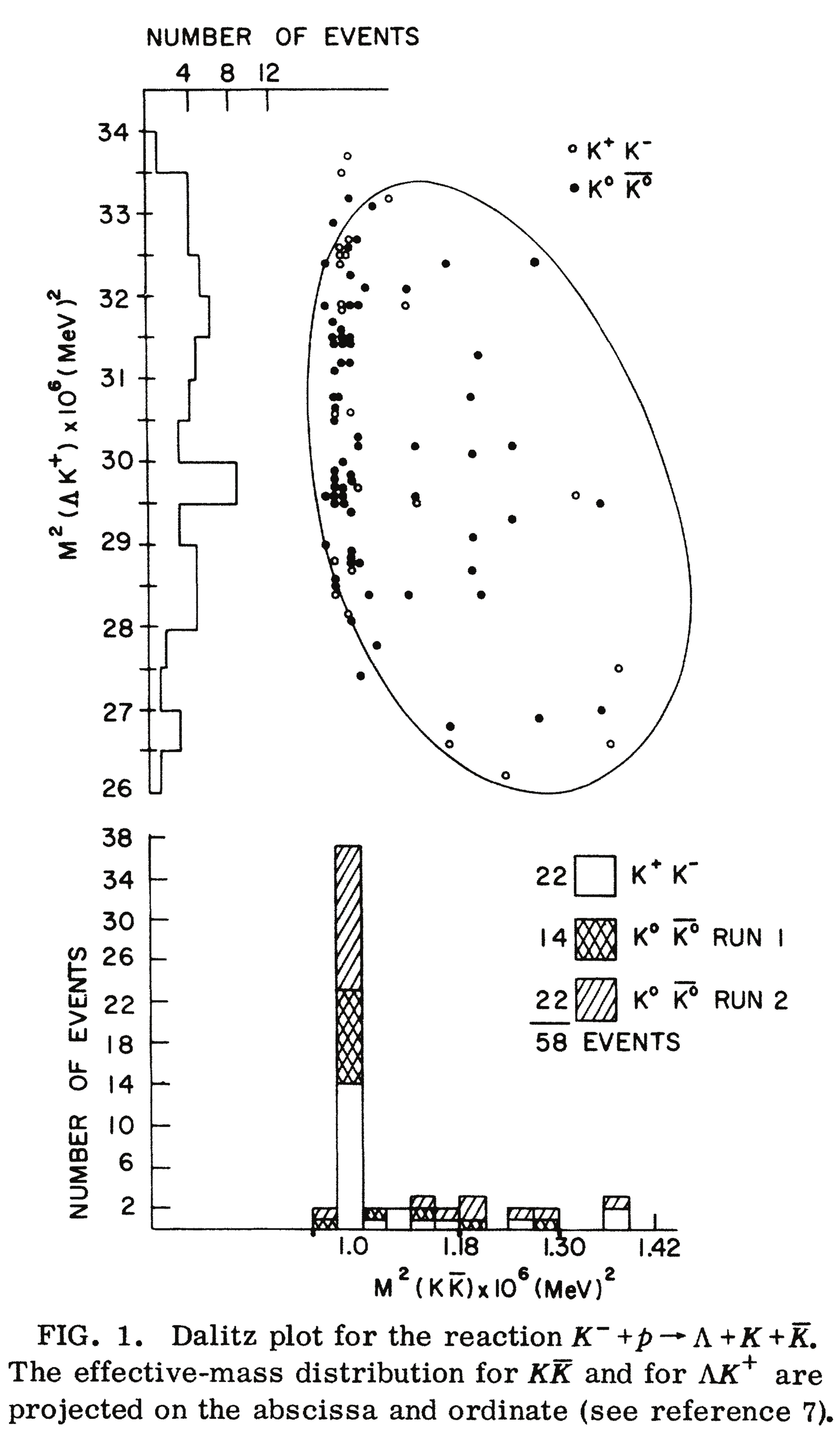}
\caption{Dalitz plot taken from Ref.~\cite{phi2}. The expected dominant decay, $\phi\to \rho+\pi$, was not observed. Instead, $\phi$ decayed into $K+\bar K$, even though the {\it K} and $\bar K$ have angular momentum 1, and
all resonant events are at the edge of the Dalitz plot. Reprinted with permission. Copyright 1963 by the American Physical Society.
}
\label{fig:dalitz}
\end{figure}

{\it How constituent quarks (aces) were discovered}~\cite{GZorigins}: Let me tell you about just one of those meetings, which took place late April 1963. On April 15, {\it Physical Review Letters} published a paper titled ``Existence and Properties of the $\phi$ Meson''~\cite{phi2}. The casual reader of that article, and perhaps even the authors themselves, might have thought this was just a confirmation of the existence of yet another resonance. By then over 25 ``credible'' meson resonances had been reported.
But I thought it remarkable that the $\phi$ decayed only into $K + \bar K$ near threshold, with angular momentum 1, while there was no evidence for the decay into $\rho + \pi$ far above threshold, with angular momentum 0. Phase space arguments greatly favored $\rho + \pi$ over $K + \bar K$, but only $K + \bar K$ was observed. My calculations showed that the decay into $\rho +\pi$ was suppressed by at least two orders of magnitude. The $\phi$ was much narrower than expected (see Fig.~\ref{fig:dalitz})!

How was this discrepancy to be understood? The authors of the paper noted that there might be a problem, but dismissed the discrepancy. They wrote:
\begin{quote}
``The observed rate [for $\phi\to\rho+\pi$]  is lower than ... predicted values by one order of magnitude; however the above estimates are uncertain by at least this amount so that this discrepancy need not be disconcerting.'' 
\end{quote}

Feynman couldn't be bothered with the discrepancy. He launched into a tirade about how unreliable experiments were, and explained that at the time he proposed the V--A theory for the weak interactions,  experiments were against him, and those experiments all turned out to be wrong.\footnote{The V--A theory was initially at variance with angular correlations measured in He${^{6}}$ decay, and the absence of the decay $\pi^{-} \to e^{-}+\bar\nu$. Later at CERN, Alvin observed this decay at the predicted rate, confirming V--A.}

But I couldn't get the suppression of $\phi$ decay out of my mind. Feynman had taught that ``in the strong interactions everything that can possibly happen does, and with the maximum strength allowed by unitarity.''\footnote{This was a different, but more useful, form of Murray's Totalitarian Principle:  ``Everything which is not forbidden is compulsory.''}  Well here was a strong interaction --- a decay --- that was not happening with maximal strength.  It wasn't happening at all! Current theory said that suppressions exist because of symmetries, but in this case there wasn't a symmetry to enforce the suppression. I was convinced that something important must be happening.

In 1949 Fermi and Yang suggested that the pion was not an elementary particle, but rather a bound state of a nucleon and antinucleon~\cite{Fermi}. Sakata extended that model to include strangeness, using $p$, $n$, and $\Lambda$ to form both meson and baryon resonances. By 1963 enough was known about hadron dynamics and the baryon resonances to see that these models could not be correct in detail,\footnote {Indeed, Fermi and Yang had written "Unfortunately we have not succeeded in working out a satisfactory relativistically invariant theory of nucleons among which ... attractive forces act [to form pions]."} but the idea that hadrons had constituents fascinated me. I replaced Sakata's constituents with three unknown constituents, $p_{0}$, $n_{0}$, and $\Lambda_{0}$~\cite{GZaces1,GZaces2},
and called them ``aces.''\footnote{There are 4 aces in a deck of cards, so why call them aces? Because in analogy with the 4 leptons known at that time, I though that there should be a fourth constituent. If the $\tau$ were known then, I might have called them dice.}
The first two aces had strangeness 0, the third, $\Lambda_{0}$, strangeness --1. To avoid problems with the baryon spectrum inherent in the Sakata model, aces were assigned baryon number 1/3. Fractional baryon number meant fractional charge.  The mass splitting between the  $p_{0}$ and $n_{0}$ was assumed to be of electromagnetic origin, and therefore small. The $\Lambda_{0}$ was assumed to be substantially heavier than the other two aces, and responsible for the SU(3) symmetry breaking that occurred in the strong interactions.
The $\phi$ was assumed to consist entirely of $\Lambda_{0}\bar\Lambda_{0}$, and the $\rho$ and the $\pi$ to consist only of the {\it other} two aces and their antiparticles. I didn't want the $\phi$ to contain any $p_0\bar p_0$ or $n_0\bar n_0$, since the strong interactions distinguished $\Lambda_{0}$ from $p_0$ and $n_0$. Assuming that the squares of meson masses were proportional to the sum of the squares of the masses of their constituents led to two relations among vector meson masses,
$$
m_{\omega}^{2}=m_{\rho}^{2},
$$
\vskip -0.2in
and
\vskip -0.2in
$$
m_{K^{*}}^{2} =  ( m_{\phi}^{2}+ m_{\rho}^{2})/2.
$$
Both relations were remarkably accurate.

\begin{figure}
   \centering
   \includegraphics [scale=.5] {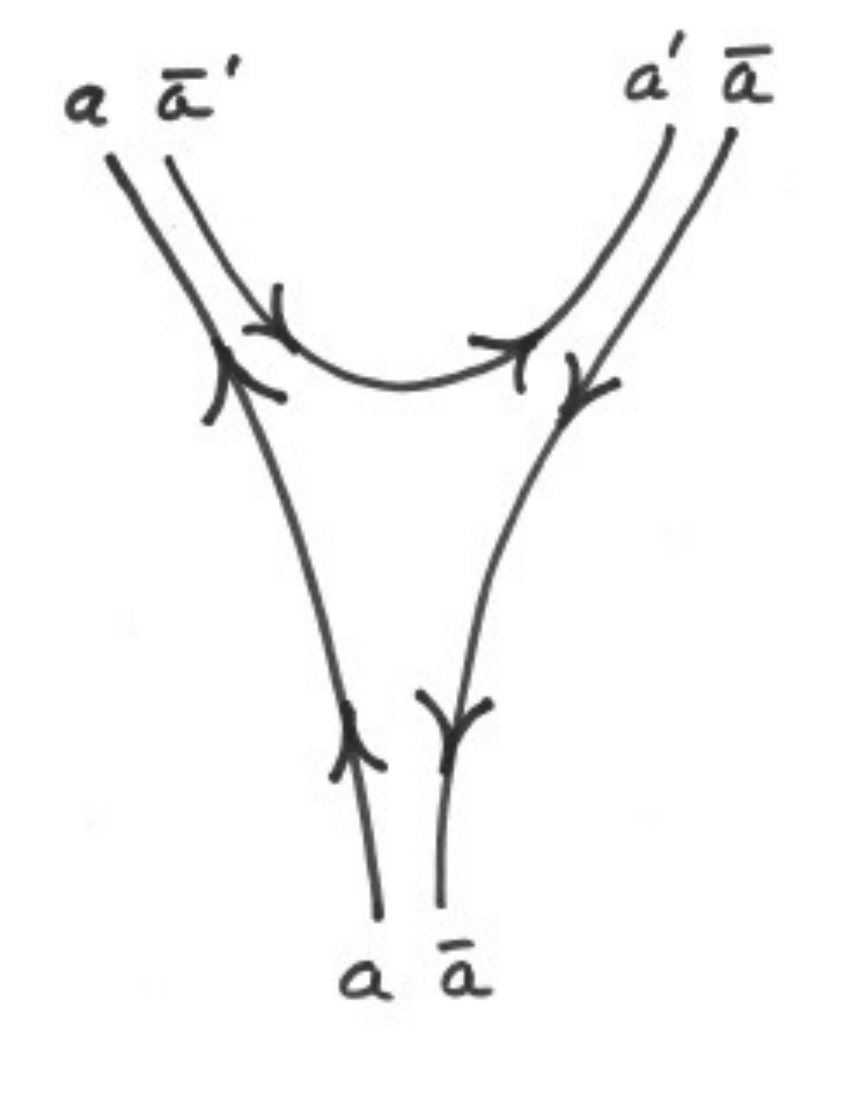}
   \caption{``Zweig diagram'' for the decay of the meson $a\,\bar a$. Murray sometimes called these ``twig diagrams,'' since the English word ``twig'' is derived from the German word, ``zweig,'' meaning branch.
   }
   \label{fig:branchGraph}
\end{figure}

What remained was an assumption about dynamics, i.e., an assumption about how mesons decay, expressed in terms of their constituents. I assumed that when a meson $a\bar a$ initiated its decay into two other mesons $a\bar a^{'} + a^{'}\bar a$, the $a$ would separate from the $\bar a$, and as the separation increased, a new $a^{'}\bar a^{'}$  pair would pop out of the vacuum, also separate, and combine with the now separated $a$-$\bar a$ pair to complete the decay (see Fig.~\ref{fig:branchGraph}),\footnote{The $a$ and $\bar a$ were not allowed to separate without the creation of an $a^\prime \bar a^\prime$ pair, since aces had fractional charge, and fractionally charged particles were not observed in meson decays. The other possibility, that the $a$ and $\bar a$ would ``eat each other,'' was forbidden by fiat.}
$$
a\bar a \rightarrow a\bar a^{'} + a^{'}\bar a.
$$
Since the $\phi$ only contained $\Lambda_0$ and $\bar\Lambda_0$, whereas $\rho$ and $\pi$ only contained $n_0,\ p_0,\ \bar n_0,$ and $\bar p_0$,  $\phi$ decay into $\rho + \pi$ was impossible!

\begin{figure}
   \centering
   \includegraphics [scale=.5] {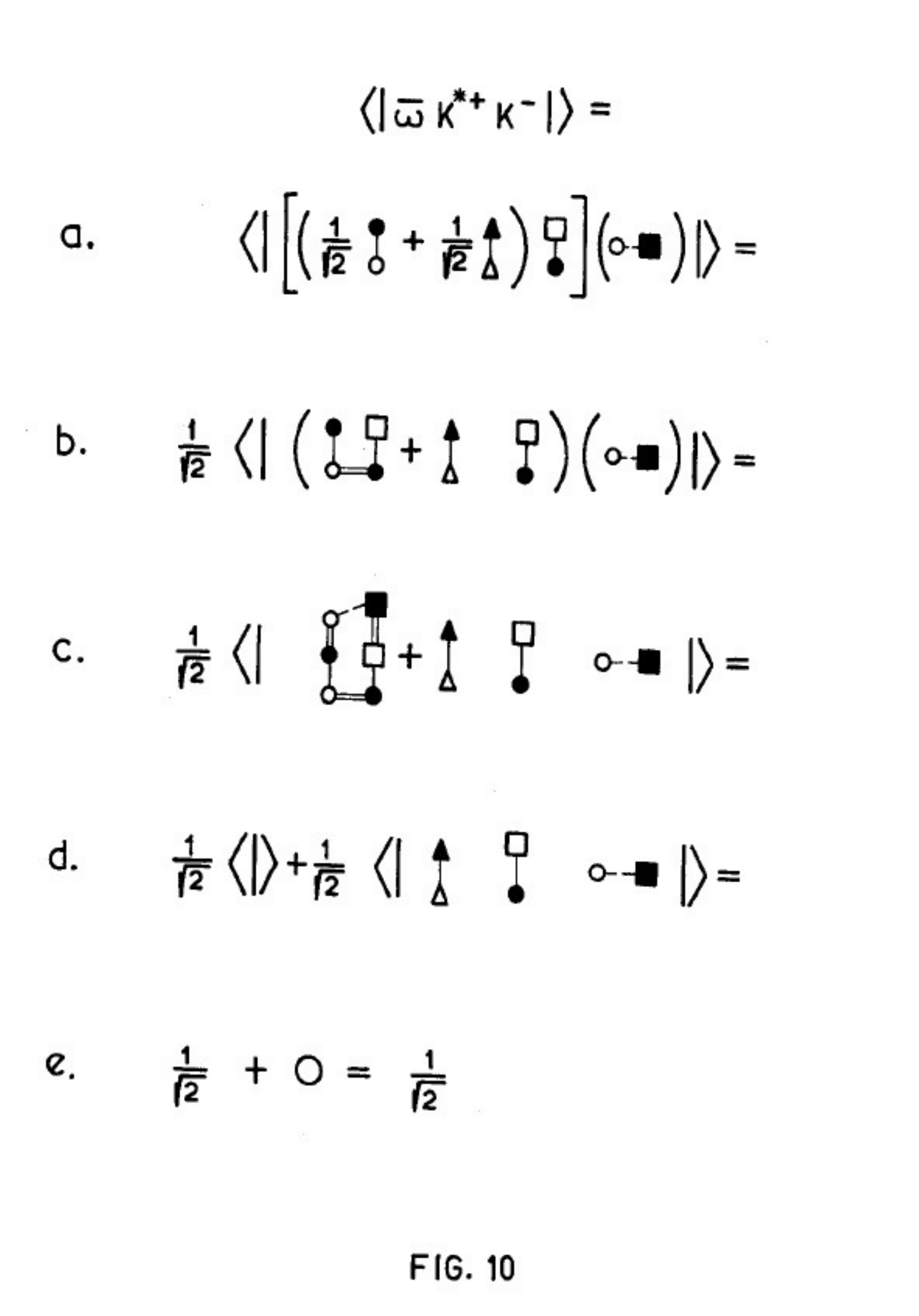}
   \caption{The graphical computation of the $\omega K^{*+} K^{-}$ coupling constant taken from the original ace paper~\cite{GZaces2}. Circles, triangles, and squares represent $p_0,\ n_0,\ {\rm and}\ \Lambda_0$, respectively; antiaces are shaded. A meson is formed by tying an ace to an antiace with a string (straight line). Strong interaction symmetry is broken by making the $\Lambda_0$ heavier (larger) than the other two aces. Additional aces, if discovered, were to be represented by pentagons, hexagons, etc. This idiosyncratic graphical calculus did not facilitate the acceptance of aces as constituents of hadrons.}
   \label{fig:aceFig10}
\end{figure}

The amplitude for any hadronic decay could be computed pictorially. Fig.~\ref{fig:aceFig10} is an example taken from the original ace paper~\cite{GZaces2}. These diagrams contained more information than SU(3) provided. ``Zweig's rule'' not only forbad certain decays, it specified the relative amplitudes of allowed decays.\footnote{Explicit rules for computing decay amplitudes implicit in the graphical calculus are summarized in Appendix 2 of Ref.~\cite{MWZpatterns}.} For example, in addition to forbidding $\phi\to\rho+\pi$, the rule determined the F/D ratio for meson-baryon couplings.

{\it Were aces real?} Since aces obeyed dynamical rules, it was hard to imagine that they weren't real. Ace-antiace pairs popped out of the vacuum in hadronic decays. Aces and antiaces orbited around one another with angular momentum $\vec L$ and total spin $\vec S$; the mesons they created had mass that depended on the value of $\vec L\cdot\vec S$.
And the weak leptonic decay of hadrons was attributed to the weak decay of their ace constituents, which were governed by V--A interactions. However, arguing against the reality of aces was the existence of the famous spin $3\over2$ $\Omega^-$, which contained 3 identical $\Lambda_0$ aces with their spins aligned, violating Pauli's spin-statistics theorem!~\footnote{I thought this problem would eventually be solved, and it was, by distinguishing the 3 aces with 3 different colors.}

This ``tinker-toy'' view of hadron physics that seemed to violate the spin-statistics theorem drove people crazy.\footnote{In addition, since aces hadn't been observed, doing physics with aces ignored a fundamental lesson learned from quantum mechanics: ``Always work with observables.'' The ``Bootstrap,'' built on Heisenberg's {\it S} matrix of scattering amplitudes, evolved from this maxim.} When I went in to see Murray to explain my ideas after returning from CERN in the early fall of 1964, he exclaimed ``Oh, the concrete quark model. That's for blockheads!'' When I explained my reason for the suppression of $\phi$ decay to Feynman, he became visibly irritated, arguing that ``unitarity mixes all states with the same quantum numbers,'' making suppression impossible. For example, the $\phi$ mixes with the $\omega$, which mixes with the $\rho+\pi$, so that $\phi$ must go to $\rho+\pi$. I was saying that the $\omega$ and $\phi$ mix, but in just such a way as to make the $\phi$ consist entirely of $\Lambda_{0}\bar\Lambda_{0}$, forbidding the decay into $\rho+\pi$. It might seem to have been a bizarre assumption, but I had no alternative.\footnote{Even today, knowing about QCD, the suppression of $\phi$ into $\rho + \pi$ is still somewhat mysterious.} It wasn't until more than a decade later, with the discovery of the exceptionally narrow $\psi$/J, that people realized that the $\phi$ and the $\psi$/J were narrow for similar reasons, and finally accepted the idea that hadrons have constituents with dynamics that obey Zweig's rule.

{\it Murray's toy field theories:}
Murray had a completely different view of quarks, using them as fundamental fields in a toy field theory. Murray's use of field theories, from which symmetry relations could be abstracted, first appeared in a 1957 article~\cite{MGMmodel}. The abstract begins with:
\begin{quote}
``An attempt is made to construct a crude field theory of hyperons and K particles, which are assumed to
have spin 1/2 and spin 0, respectively.''
\end{quote}
Fields in this model correspond to real particles, e.g., the $\Lambda$ and {\it K}, and Murray establishes relations between meson-baryon coupling constants by assuming global symmetry. Most enlightening, however, are the ``General Remarks:''
\begin{quote}
``Supposing that the model we have presented has elements of truth, we may add the following remarks:
\vskip 0.005in
(1) The symmetry properties of the model may be correct even though the use of field theory is unjustified. For this reason an analysis purely in terms of the symmetry group of the theory is in order."
\end{quote}
Here Murray constructs a field theory that he knows is incorrect in detail, picks properties of the objects in the theory that he believes should also hold in the real theory, and throws away the rest.

Four years later in the ``Eightfold Way,'' Murray proposes that unitary symmetry be used to classify particles, rather than global symmetry, this time using hypothetical particles $l$ and $\bar L$ as fundamental fields~\cite{MGMeight}:
\begin{quote}
``For the sake of a simple exposition, we begin our discussion of unitary symmetry with `leptons' [$l$ and $\bar L$], although our theory really concerns the baryons and mesons and the strong interactions. The particles we consider here for mathematical purposes do not necessarily have anything to do with real leptons, but there are some suggestive parallels.''
\end{quote}

After using $l$ and $\bar L$ to construct states that transform like real particles, Murray reassures the reader that:
\begin{quote}
``We shall attach no physical significance to the $l$ and $\bar L$ `particles' out of which we have constructed the baryons. The discussion up to this point is really just a mathematical introduction to the properties of unitary spin.''
\end{quote}

The ``Eightfold Way'' was never published in a journal. Ideas from it were distilled, leading to a much more formal paper with a toy field theory based on the Sakata model, and not the model based on $l$ and $\bar L$~\cite{MGMsymmetries}. From Section IV of that paper:
\begin{quote}
``We generalize the Fermi-Yang description to obtain the symmetrical Sakata model and abstract from it as many physically meaningful relations as possible.''
\end{quote}

{\it Current quarks, 1964:} 
According to Bob Serber, in the spring of 1963 over lunch at the Columbia faculty club, Serber told Murray about a scheme he had been thinking about in which baryon representations were made from three fundamental representations of SU(3) ($3\times3\times3$), and meson representations from the fundamental representation and the representation representing the antiparticles of the fundamental representation ($3\times\bar3$).\footnote{Letter to me from Bob Serber dated July 8, 1980.} After a moment's calculation Murray found that this would imply that the members of the fundamental representation would have fractional charge, a fact that Serber had not realized. No more was said, but in February of 1964 Murray proposed using the three fractionally charged objects in the fundamental representation as fields from which to construct the currents of a toy field theory~\cite{MGMschematic}.
\begin{quote}
``we assign to the triplet $t$ the following properties: spin $1\over2$, $z = -{1\over 3}$, and baryon number $1\over 3$.  We then refer to the members $u^{2\over3}$, $d^{-{1\over3}}$, and $s^{-{1\over3}}$ of the triplet as `quarks' ...
\noindent
A formal mathematical model based on field theory can be built up for the quarks exactly as for p, n, $\Lambda$ in the old Sakata model ...
\noindent
All these [current commutation] relations can now be abstracted from the field theory model and used in a dispersion theory treatment.''
\end{quote}
Finally, Murray ends the paper with the famous lines:
\begin{quote}
``It is fun to speculate about the way quarks would behave if they were physical particles of finite mass (instead of purely mathematical entities as they would be in the limit of infinite mass). ...
A search for stable quarks of charge  $-{1\over3}$ or  $+{2\over3}$ and/or stable di-quarks of charge $-{2\over3}$ or
$+{1\over3}$ or $+{4\over3}$ at the highest energy accelerators would help to reassure us of the non-existence of real quarks.''
\end{quote}

Murray's modus operandi is eloquently explained in a paper published five months later~\cite{MGMcurrents}:
\begin{quote}
``We use the method of {\it abstraction} from a Lagrangian field theory model. In other words, we construct a mathematical theory of the strongly interacting particles, which may or may not have anything to do with reality, find suitable algebraic relations that hold in the model, postulate their validity, and then throw away the model. We compare this process to a method sometimes employed in French cuisine: a piece of pheasant meat is cooked between two slices of veal, which are then discarded.''
\end{quote}

{\it Murray's evolving view of quarks:} At the end of February 1972, Murray delivered a set of lectures in Schladming Austria titled ``Quarks''~\cite{MGMquarks}. This is the last record I have showing Murray's views before the ``November Revolution'' when the $\psi/J$ was discovered, making the existence of real quarks all but obvious. By that time Murray spoke of ``constituent quarks,'' but viewed {\it his} quarks as ``current quarks.'' Murray begins with:

\begin{quote}
``In these lectures I want to speak about at least two interpretations of the concept of quarks for hadrons and the possible relations between them. First I want to talk about quarks as `constituent quarks'. These were used especially by G. Zweig (1964)
who referred to them as aces. ...
The whole idea is that hadrons act as if they are made up of quarks, but the quarks do not have to be real. If we use the quark statistics described above, we see that it would be hard to make the quarks real, since the singlet restriction is not one that can be easily applied to real underlying objects; ...
\newline
\newline
There is a second use of quarks, as so-called `current quarks' which is quite different from their use as constituent quarks; ...
In the following discussion of current quarks we attempt to write down properties that may be exact, at least to all orders in the strong interaction, with the weak, electromagnetic and gravitational interactions treated as perturbations. ...
\newline
\newline
If quarks are only fictitious there are certain defects and virtues. The main defect would be that we never experimentally discover real ones and thus will never have a quarkonics industry. The virtue is that then there are no basic constituents for hadrons --- hadrons act as if they were made up of quarks but no quarks exist --- and, therefore, there is no reason for a distinction between the quark and bootstrap picture: they can be just two different descriptions of the same system, like wave mechanics and matrix mechanics. In one case you talk about the bootstrap and when you solve the equations you get something that looks like a quark picture; in the other case you start out with quarks and discover that the dynamics is given by bootstrap dynamics. ...''\footnote{However, in order to recover bootstrap dynamics, the algebraic properties of operators abstracted from the free field theory of current quarks would have to be supplemented by additional assumptions about quark dynamics.}
\newline
\newline
``If we go too far...and try to construct a complete Fock space for quarks and antiquarks on a light-like plane, abstracting the algebraic properties from free quark-theory, we are in danger of ending up with real quarks, and perhaps even with free real quarks as I mentioned before. In our work, we are always between Scylla and Charybdis; we may fail to abstract enough, and miss important physics, or we may abstract too much and end up with fictitious objects in our models turning into real monsters that devour us.''
\end{quote}

The 1957 article ``Elementary Particles,'' that Murray wrote with Rosenbaum,  was viewed as a great success by Jim Flanagan, editor of the {\it Scientific American}. In late 1971 he flew Frank Bello, an associate editor, to Pasadena to help Murray and me write an article on quarks, the new ``elementary particles.''  Frank and I wrote a draft, but he and Murray completely rewrote it after Frank got back to New York, changing the meaning of constituent quarks. Murray and Frank wrote:\footnote{From a draft Frank sent to Murray on February 25, 1972; Caltech Archives.}

\begin{quote}
``As seemed probable from the outset, the quark model may be nothing more than a useful mathematical construct: The known hadrons --- including dozens not yet discovered when the model was conceived --- behave `as if' they were composed of quarks. Quarks themselves may have no independent existence.''
\end{quote}
Murray and I could not agree on the meaning of constituent quarks.\footnote{Valentine Telegdi provides an independent description of our differing views~\cite{Telegdi}.} When Murray suggested we abandon the article, I agreed.

\begin{figure}
   \centering
   \includegraphics[scale=0.8]{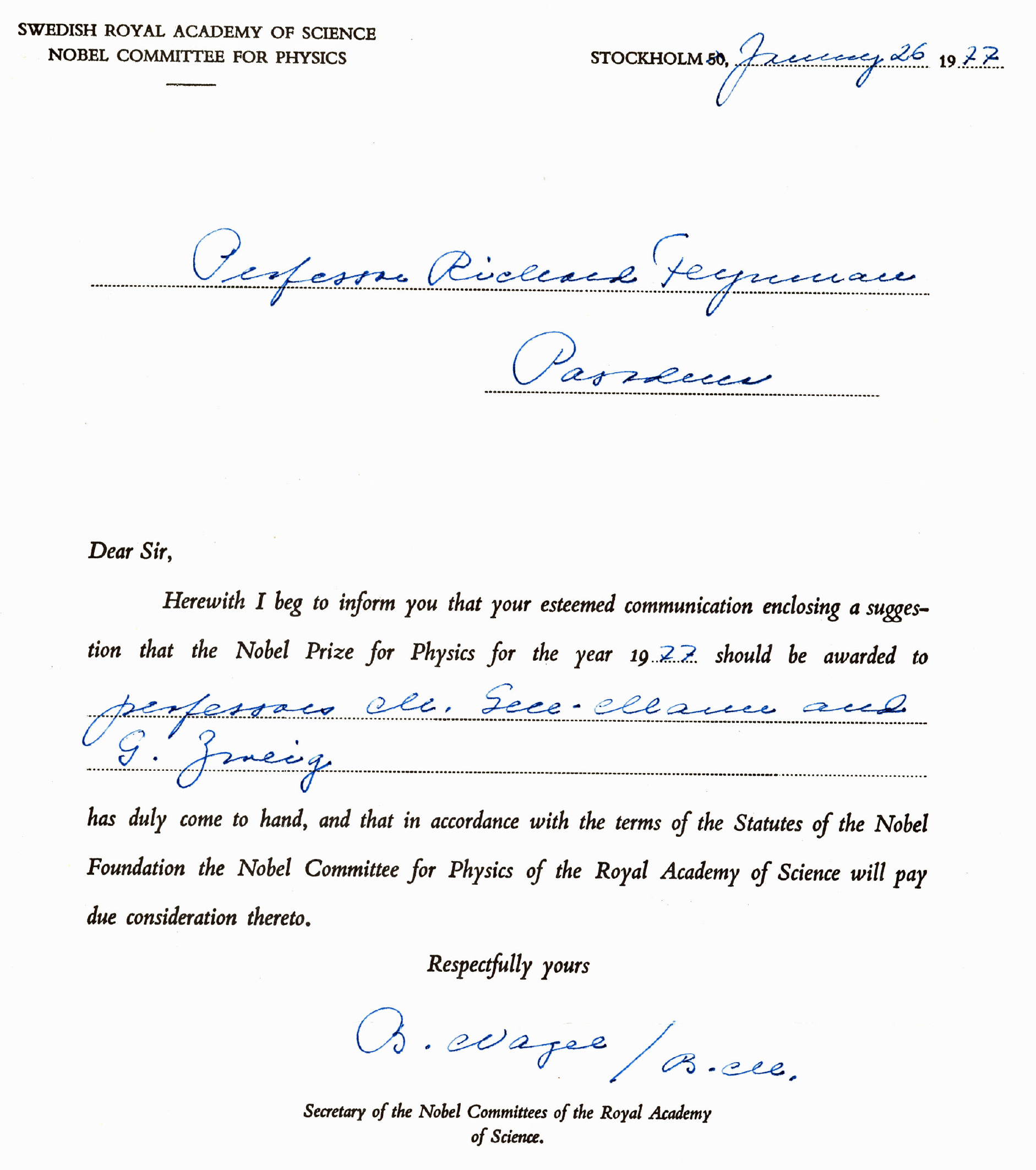}
   \caption{Acknowledgement of Feynman's nomination letter by the Nobel Committee for Physics.}
   \label{fig:letter}
\end{figure}

{\it A tribute from the master:}
In 1977 Feynman nominated both of us for the Nobel Prize in Physics. When I learned about this relatively recently, I felt great satisfaction. Murray, on the other hand, might think that this is no big deal for him. After all, he already has a Nobel prize, and he presumably gets nominated every year for a second one. But to my knowledge, Feynman never nominated anyone for anything, so I think this is a real tribute, even for Murray. As proof of Feynman's nomination, I offer Fig.~\ref{fig:letter}.\footnote{Murray was delighted to hear of Feynman's nomination. He was unaware of it before this talk.}

{\it Summary:} How can Murray's contributions, described here,  be put in some perspective? Causality and CPT symmetry are examples of very general principles that are expected to hold in all particle interactions. By abstracting from toy free field theories, Murray identified certain algebraic relations among operators, e.g., the equal-time current commutation relations, that he postulated as also holding in the strong interactions. Since the matrix elements of these operators were measurable, his postulates were testable, and some were quickly verified to reasonable accuracy~\cite{SAsumRules,WWsumRules}.  These relationships between operators, though limited in scope,\footnote{Limited by virtue of their existence in a free field theory.} could be absolutely true. His genius was to understand that he must find Scylla and Charybdis, and then, like Jason, sail between them. 

Science is a social enterprise, and society recognizes individuals that influence the work of others. Murray was concerned with describing reality, making predictions that could be tested experimentally, and providing a theoretical framework that enabled others to expand on his vision. The reason we are here today is because Murray thereby set an agenda for an entire generation of physicists, dominating our field like no other.

{\it Epilogue:} Murray's work eventually became less concerned with experiment, and more with theory. I walked into his office one day and asked, ``Murray, you're so good at phenomenology, why aren't you doing it?'' He replied, ``I'm not interested in it any more.'' I was shocked. It was like Picasso in his prime giving up on painting. It was the end of an era.

Over the years Murray and I drifted apart.
Murray worked on foundations of quantum theory, then complexity and linguistics.  I switched to neurobiology, or as Murray put it with a smile, ``cutting up cats.''

The historian Dan Kevles --- whose office I barged into almost 50 years earlier, asking him to record history in the making --- true to his word, went on to research the past and write about George Ellery Hale in the Gilded Age, and Robert Millikan, a founder and first president of Caltech.
Eventually Dan did broaden his vision of what historians do. In 1998 he wrote a book about Caltech's then sitting president David Baltimore~\cite{DKbaltimore}.

Oh, and whatever happened to aces? They are alive and well! In case you haven't noticed,  constituent quarks are really aces in disguise.

{\it Acknowledgements:} Charlotte Erwin, Head of Archives and Special Collections at Caltech, and Loma Karklins have been very helpful in locating documents for this talk. Erica Jen has provided invaluable advice as to what to say, and how to say it.

\end{document}